%% file: maxminSINR_stanczak.tex
\documentclass[12pt,draftclsnofoot,onecolumn]{IEEEtran}

\input{makros}
\usepackage{enumerate}
\usepackage[nowinedt]{srcltx}

\title{A Characterization of Max-Min SIR-Balanced Power Allocation with
  Applications}

\author{
S\l awomir Sta\'nczak$^{\dag}$, Micha\l\ Kaliszan$^{\dag}$, Nicholas Bambos$^{\ast}$ and 
Marcin Wiczanowski$^{\dag}$\\
	{\centering
		$^{\dag}$Fraunhofer German-Sino Lab for Mobile Communications\\
    Einsteinufer 37,  D-10587 Berlin, Germany\\
    Email: \{stanczak, michal.kaliszan, marcin.wiczanowski\}@hhi.fraunhofer.de\\
		$^{\ast}$Department of Electrical Engineering\\
    350 Serra Mall, Stanford University, Stanford, CA 94305\\
    Email: bambos@stanford.edu
	}
}

\date{}

\begin{document}
\maketitle
\begin{abstract}
  We consider a power-controlled wireless network with an established network
  topology in which the communication links (transmitter-receiver pairs) are
  corrupted by the co-channel interference and background noise. We have
  fairly general power constraints since the vector of transmit powers is
  confined to belong to an arbitrary convex polytope. The interference is
  completely determined by a so-called gain matrix.  Assuming irreducibility
  of this gain matrix, we provide an elegant characterization of the max-min
  SIR-balanced power allocation under such general power constraints. This
  characterization gives rise to two types of algorithms for computing the
  max-min SIR-balanced power allocation. One of the algorithms is a
  utility-based power control algorithm to maximize a weighted sum of the
  utilities of the link SIRs. Our results show how to choose the weight vector
  and utility function so that the utility-based solution is equal to the
  solution of the max-min SIR-balancing problem. The algorithm is not amenable
  to distributed implementation as the weights are global variables. In order
  to mitigate the problem of computing the weight vector in distributed
  wireless networks, we point out a saddle point characterization of the
  Perron root of some extended gain matrices and discuss how this
  characterization can be used in the design of algorithms in which each link
  iteratively updates its weight vector in parallel to the power control
  recursion. Finally, the paper provides a basis for the development of
  distributed power control and beamforming algorithms to find a global
  solution of the max-min SIR-balancing problem.
\end{abstract}

\begin{IEEEkeywords}
Max-min SIR-balancing, Max-min fairness, Power control, 
Wireless networks, Utility maximization, Interference management, 
Distributed algorithms
\end{IEEEkeywords}


\section{Introduction}
\label{sec:intro}

Wireless channel is error-prone and highly unreliable being subject to several
impairment factors that are of transient nature, such as those caused by
co-channel interference or
multipaths. Excessive interference can significantly deteriorate the network
performance and waste scarce communication resources.  For this reason,
strategies for resource allocation and interference management are usually
necessary in wireless networks to provide acceptable QoS
levels to the users.

There are different mechanisms for resource allocation and interference
management. Power control may play a central role in distributed wireless mesh
networks, where, due to the lack of a central network controller, link
scheduling strategies are notoriously difficult to implement. Thus, a
reasonable approach is to avoid only strong interference from neighboring
links, and then use an appropriate power control policy to manage the
remaining interference in a network. In this paper, we focus on the power
control problem, which addresses the issue of coordinating transmit powers of
links such that the worst signal-to-interference ratio (SIR) balanced against
some SIR targets attains its maximum. This so-called max-min SIR-balancing
problem is a widely studied resource allocation problem for wireless networks
(see, for instance, \cite
{aein73,meyerhoff74,alavi82,Zander92,Zander92a,Foschini93,Yates95a,Yates95b,Bambos98,Elbatt04,zander01,YangXu98}
as well as \cite[Sections 3.1 and 5.6]{SchubBoche06NoW},
\cite{BocheSchubertSignalProc06}, \cite[Section 5.6]{StBookSpringer08}). A key
feature of this strategy is that any given SIR (signal-to-interference ratio)
targets are feasible if and only if they are satisfied under a max-min
SIR-balanced power allocation.  Moreover, the notion of max-min SIR-balancing
is closely related to max-min fairness, the most common notion of
fairness. Note that since we focus on static wireless networks, transmit
powers are to be periodically adjusted to changing channel and network
conditions (dynamic power control). This in turn presumes a relatively low up
to moderate network dynamics. In contrast, in highly dynamic wireless
networks, one should consider resource allocation schemes for
\emph{stochastic} wireless networks
\cite{KanduBoyd02,PapaEvansDey05,PapaEvansDey06,BorstWhiting03,LeeMazumdar06,NeelyThesis,NeelyModInfoCom2005,StolyarQS05}.

In the \emph{noiseless} case, which is widely considered in the literature
\cite{aein73,meyerhoff74,alavi82,Zander92a,Zander92,gerlach96,he97,hongyu00}.
(an overview can be found in \cite{zander01,montalbano98}) and where power
constraints play no role in the analysis, it is widely known \cite{zander01},
\cite[Sections 3.1]{SchubBoche06NoW} that any positive eigenvector of the
(irreducible) gain matrix scaled by a diagonal matrix of given SIR targets is
a solution to the max-min SIR-balancing problem. In \cite{YangXu98}, the
problem was solved for a ``noisy'' downlink channel constrained on total
power. The sum constraint on transmit powers was captured by an additional
equation so that the optimal solution is characterized in terms of some unique
eigenvector of a certain irreducible gain matrix of higher dimension (see also
\cite[pp. 111-113]{SchubBoche06NoW}). 

Assuming an irreducible gain matrix, Sections
\ref{sec:UserPC_MaxMin_Uniqueness}--\ref{sec:UserPC_MaxMin_IndCon} extend
these results to any convex polytope as the constraint set to model
constraints on transmit powers of the links. In addition to the analysis in a
higher dimension (as in \cite{YangXu98}), we also obtain an elegant
characterization of the max-min SIR-balanced power vector from an eigenvalue
problem of the same dimension as the original problem. These results were
inspired by \cite{MahdaviISIT07}, where the authors used different tools to
characterize rate region in interference channel with constrained power (see
also the acknowledgments after the conclusions at the end of this paper).

In Section \ref{sec:applications}, we use the results of Sections
\ref{sec:UserPC_MaxMin_Uniqueness}--\ref{sec:UserPC_MaxMin_IndCon} to
establish a connection between the max-min SIR-balancing power control problem
and the utility-based power control problem. Such a connection is known in the
noiseless case \cite[Section 5.9]{StBookSpringer08} and constitutes the
starting point for the analysis in \cite{BocheSchubertSignalProc06}. More
precisely, we show how to choose the weight vector and utility function so
that solving the problem of maximizing a weighted sum of the utilities of the
SIRs leads to the max-min SIR-balancing solution. This result was used in
\cite{FeistelStKalWSA09} to solve the max-min SIR-balancing problem over the
\emph{joint} space of admissible power vectors and receive beamformers. Thus,
the results of this paper provide key tools to generalize the results of
\cite{BocheSchubertSignalProc06} to noisy channels under general power
constraints and to a larger class of utility functions.

An advantage of the utility-based approach is that there exist distributed
power control schemes to compute the max-min SIR power vector, provided that
each link knows how to select its weight
\cite{StWiBoPowerControlJour06,HandeChiang08}. The problem is however that a
desired weight vector is determined by positive left and right eigenvectors of
some nonnegative matrix, so that the links cannot choose their weights
independently. Thus, as neither the eigenvectors nor the corresponding
matrices are a priori known at any node, the presented approach for computing
the max-min SIR power allocation is still not amenable to implementation in
decentralized wireless networks.

A basic idea to overcome or at least to alleviate this problem is to let each
link iteratively update its weight vector in parallel to the power control
recursion. In Section \ref{sec:appl_saddlePC}, we point out that a saddle
point characterization of the Perron roots of some nonnegative matrices. This
characterization provides a basis for efficient saddle point algorithms
converging to a max-min SIR-balanced power vector. Basically, the idea is
redolent of primal-dual algorithms that employ some optimization methods to
minimize the Lagrangian over primal variables and to \emph{simultaneously}
maximize it over dual variables. 

Before starting with the analysis, we introduce definitions, notation and
state the max-min SIR-balancing problem. 

\section{Definitions and problem statement}
\label{sec:model}

We consider a wireless network with an established network topology, in which
all links share a common wireless spectrum.  Let $K\geq 2$ users (logical
links) compete for access to the wireless links and let $\logic=\{1,\dotsc,
K\}$ denote the set of all users. The transmit powers $p_k,k\in\logic,$ of the
users are collected in the vector $\ve{p}=(p_1,\dotsc,p_K)\geq 0$, which is
referred to as \emph{power vector} or \emph{power allocation}. The transmit
powers are subject to power constraints so that $\ve{p}\in\pset$
where\footnote{$\RN,\RP$ are nonnegative and positive reals, respectively.}
\begin{equation}
  \label{eq:NetMod_PowerConstrSetMatrixForm}
\pset=\{\ve{p}\in\RN^K: \ma{C}\ve{p}\leq\hat{\ve{p}},\ma{C}\in\{0,1\}^{N\times K}\}\subset\R^K
\end{equation}
for some given $\hat{\ve{p}}=(P_1,\dotsc,P_N)>0$ and $\ma{C}$ with at least
one $1$ in each column so that $\pset$ is a compact set. Throughout the
paper, we use $\nodes=\{1,\dotsc,N\}$ where $N$ is the number of power
constraints. The main figure of merit is the SIR at the output of each
receiver given by
\begin{enumerate}[\theassume]
\refstepcounter{assume}\label{as:InterferenceLinear}
\item $\sir_k(\ve{p}) =p_k/I_k(\ve{p}),k\in\logic,$ where the interference
  function $I_k$ is $I_k(\ve{p})=(\vmat\ve{p}+\ve{z})_k=\sum_{l=1}^K
  v_{k,l}p_l+z_k$.
\end{enumerate}
$\vmat:=(v_{k,l})\in\RN^{K\times K}$ is the \emph{gain matrix},
$v_{k,l}=V_{k,l}/V_{k,k}$ if $l\neq k$ and $0$ if $l=k$ where $V_{k,l}\geq 0$
with $V_{k,k}>0$ is the attenuation of the power from transmitter $l$ to
receiver $k$. The $k$th entry of $\ve{z}:=(z_1,\dotsc,z_K)$ is
$z_k=\sigma_k^2/V_{k,k}$, where $\sigma_k^2>0$ is the noise variance at the
receiver output.

Let $\gamma_1,\dotsc,\gamma_K>0$ be the \emph{SIR targets} and let
$\gmat=\diag(\gamma_1,\dotsc,\gamma_K)$. We say that the SIR targets are
feasible if there exists a power vector $\ve{p}\in\pset$ (called a valid power
vector) such that $\sir_k(\ve{p})\geq\gamma_k>0$.

\begin{definition}
\label{def:UserPC_MaxMin_SIRBalancedDef}
Given any $\gmat$, $\bar{\ve{p}}$ is said to be a max-min SIR-balanced power
vector if
\begin{equation}
\label{eq:UserPC_MaxMin_SIRBalancedDef}
\bar{\ve{p}}
:=\arg\underset{\ve{p}\in\pset}{\max}\min_{k\in\logic}\,(\sir_k(\ve{p})/\gamma_k)\,.
\end{equation}
\end{definition}

It is important to notice that the maximum in
(\ref{eq:UserPC_MaxMin_SIRBalancedDef}) exists as
$\min_{k\in\logic}(\sir_k(\ve{p})/\gamma_k)$ is continuous on the compact set
$\pset$. Thus, $\gamma_k$ are not necessarily met under optimal power control
(\ref{eq:UserPC_MaxMin_SIRBalancedDef}). For this reason, $\gamma_k$ can also
be interpreted as a \emph{desired} SIR value of link $k$. A trivial but
important observation is that $\bar{\ve{p}}>0$, allowing us to focus on
$\ppset=\pset\cap\RP^K$.

\section{Some Preliminary Observations}
\label{sec:UserPC_MaxMin_Uniqueness}

In general, $\bar{\ve{p}}$ of Definition
\ref{def:UserPC_MaxMin_SIRBalancedDef} is not unique. For general power
constraints, the uniqueness is ensured if $\vmat$ is irreducible
\cite{Ho85,Me00} since then the links are mutually dependent through the
interference.  In order to see this, notice that the problem
(\ref{eq:UserPC_MaxMin_SIRBalancedDef}) is equivalent to finding the largest
positive threshold $t$ such that $t\leq\sir_k(\ve{p})/\gamma_k$ for all
$k\in\logic$ and $\ve{p}\in\pset$. The constraints can be equivalently written
in a matrix form as $\gmat\ve{z}\leq(\frac{1}{t}\ma{I}-\gmat\vmat)\ve{p}$ and
$\ve{p}\in\pset$. So, as $\ve{p}$ must be a positive vector, \cite[Theorem
A.51]{StBookSpringer08} implies that the threshold $t$ must satisfy
\begin{equation}
\label{eq:UserPC_MaxMin_rho<1/t}
\rho(\gmat\vmat)<1/t\,.
\end{equation}
Now, one particular solution to the max-min SIR-balancing problem
(\ref{eq:UserPC_MaxMin_SIRBalancedDef}) is $\bar{\ve{p}}'$ given by
\begin{equation}
\label{eq:UserPC_MaxMin_SIRBalancedSpecial}
\bar{\ve{p}}'=\bigl(1/t'\ma{I}-\gmat\vmat\bigr)^{-1}\gmat\ve{z}, 
\bar{\ve{p}}'\in\pset
\end{equation}
where
\begin{align}
\label{eq:UserPC_MaxMin_tprim}
t'={\arg\max}_{t\geq 0}\, t
&&\text{s.t.}&&
(1/t\ma{I}-\gmat\vmat)\ve{p}=\gmat\ve{z}, \ve{p}\in\pset\,.
\end{align}
Note that $\bar{\ve{p}}'$ is a max-min SIR-balanced power vector such that
$\sir_k(\bar{\ve{p}}')/\gamma_k=\sir_l(\bar{\ve{p}}')/\gamma_l$ for each
$k,l\in\logic$.  This immediately follows from
(\ref{eq:UserPC_MaxMin_SIRBalancedSpecial}) when it is written as a system of
$K$ SIR equations. By (\ref{eq:UserPC_MaxMin_rho<1/t}),
(\ref{eq:UserPC_MaxMin_SIRBalancedSpecial}), (\ref{eq:UserPC_MaxMin_tprim})
and \cite[Theorem A.51]{StBookSpringer08}, $\bar{\ve{p}}'>0$ exists and is a
unique power vector corresponding to a point in the feasible SIR
region\footnote{The feasible SIR region is the subset of $\RN^K$ of all SIR
  levels that can be achieved by means of power control. $\freg$ defined by
  (\ref{eq:UserPC_MaxMin_FeasibleQoSRegion}) becomes the feasible SIR region
  if $\phi(x)=x$ and $\gamma_k=1,k\in\logic$.}  that is farthest from the
origin in a direction of the unit vector $\ve{\gamma}/\|\ve{\gamma}\|_1$.

\begin{figure}[tbp]
  \centering
    \scalebox{0.5}{\input{maxmin_problem.pstex_t}}
  \centering
  \caption{The feasible SIR region under individual power
    constraints and two different gain matrices $\vmat\geq 0$. The
    following notation is used: $\bar{\gamma}_k=\sir_k(\bar{\ve{p}})$ and
    $\bar{\gamma}_k'=\sir_k(\bar{\ve{p}}')$ where $\bar{\ve{p}}$ and
    $\bar{\ve{p}}'$ are defined by
    (\ref{eq:UserPC_MaxMin_SIRBalancedDef}) and
    (\ref{eq:UserPC_MaxMin_SIRBalancedSpecial}), respectively. \emph{Left:}
    $\vmat$ is chosen so that $\sir_2(\ve{p})=p_2/z_1$, in which case
    $\bar{\ve{p}}$ is not unique. \emph{Right:} $\vmat$ is irreducible,
    in which case $\bar{\ve{p}}$ is unique and equal to
    (\ref{eq:UserPC_MaxMin_SIRBalancedSpecial}). The point 
    $(\gamma_1'', \gamma_2'')$ corresponds to the max-min fair
    power allocation.}
\label{fig:UserPC_maxmin_problem}
\end{figure}
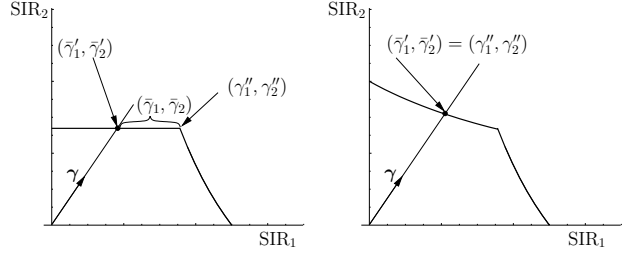

The above considerations are illustrated in Fig. \ref{fig:UserPC_maxmin_problem}.
The plots depict two examples of feasible SIR regions in a system with two users.
In both cases, the point $(\sir(\bar{\ve{p}}')_1, \sir(\bar{\ve{p}}')_2)$, 
with $\bar{\ve{p}}'$ given by \eqref{eq:UserPC_MaxMin_SIRBalancedSpecial} is the point
at the intersection of the boundary of the feasible SIR region with the line 
defined by the $\ve{\gamma}$ vector and is max-min SIR-balanced.
It is however not the unique solution if SIRs of some users can be increased 
without affecting the minimum. The examples of configurations in which 
the power allocation $\bar{\ve{p}}'$ given by \eqref{eq:UserPC_MaxMin_SIRBalancedSpecial}
is not and is a unique solution of the max-min SIR-balancing problem are presented 
in the left and in the right subplot of Fig. \ref{fig:UserPC_maxmin_problem}, respectively.

Now let us assume throughout the paper that
\begin{enumerate}[\theassume]
{\refstepcounter{assume}\label{as:UserPC_Continuous}}
\item $\phi:\RP\to\qset\subseteq\R$ is continuously differentiable and
  strictly increasing function.
\end{enumerate}
By strict monotonicity, one has
$\phi(\min_{k\in\logic}\sir_k(\ve{p})/\gamma_k)=\min_{k\in\logic}\phi(\sir_k(\ve{p})/\gamma_k)$
for every $\ve{p}>0$. Thus, as $\bar{\ve{p}}\in\ppset$,
\begin{equation}
  \label{eq:UserPC_MaxMin_SIRBalancedInFQR}
  \bar{\ve{p}}=
  \underset{\ve{p}\in\ppset}{\arg\max}\min\limits_{k\in\logic}\phi(\sir_k(\ve{p})/\gamma_k)
\end{equation}
where $\bar{\ve{p}}$ is a max-min SIR-balanced power vector defined by
(\ref{eq:UserPC_MaxMin_SIRBalancedDef}). With
(\ref{eq:UserPC_MaxMin_SIRBalancedInFQR}) in hand, we can prove a sufficient
condition for the uniqueness of $\bar{\ve{p}}$ under general power
constraints. To this end, given $\phi$, we define the set
$\menge{F}\subset\qset^K$ as
\begin{equation}
  \label{eq:UserPC_MaxMin_FeasibleQoSRegion}
  \menge{F}=\bigl\{\pv\in\qset^K:\pve_k=\phi\bigl(\sir_k(\ve{p})/\gamma_k\bigr),
  k\in\logic,\ve{p}\in\ppset\bigr\}
\end{equation}
Note that $\menge{F}$ can be interpreted as a feasible QoS region where the
QoS value for link $k$ is $\phi(\sir_k(\ve{p})/\gamma_k)$. By
\ref{as:UserPC_Continuous} and \cite[Sect. 5.3]{StBookSpringer08}, the
following can be said about $\freg$. 
\begin{observation}
  \label{obs:F_Properties}
  There is a \emph{bijective continuous map}\footnote{The reader should bear
    in mind that the existence of such a bijective map allows us to prove some
    results in $\freg$.} from $\freg$ onto $\ppset$ (see also
  (\ref{eq:bijective_map})). If
\begin{enumerate}[\theassume]
{\refstepcounter{assume}\label{as:UserPC_LogConvex}}
\item the inverse function $g(x):=\phi^{-1}(x)$ is log-convex
  \cite{StBookSpringer08},
\end{enumerate}
then $\menge{F}$ is \emph{downward comprehensive, connected and
  convex}. Moreover, $\sir_k(\bar{\ve{p}}')$ with $\bar{\ve{p}}'$ defined by
(\ref{eq:UserPC_MaxMin_SIRBalancedSpecial}) corresponds to a point on the boundary of
$\menge{F}$ where $\pve_1=\dotsc=\pve_K$ with $q_k$ defined by
(\ref{eq:UserPC_MaxMin_FeasibleQoSRegion}).
\end{observation}

Note that the boundary of $\menge{F}$ (denoted by $\partial\menge{F}$) is the
set of all points of $\menge{F}$ such that, if $\ve{p}$ is the corresponding
power vector in (\ref{eq:UserPC_MaxMin_FeasibleQoSRegion}), then
$\ma{C}\ve{p}\leq\hat{\ve{p}}$ holds with at least one equality. Widely known
examples of functions satisfying \ref{as:UserPC_Continuous} and
\ref{as:UserPC_LogConvex} are $x\mapsto\log(x),x>0,$ and $x\mapsto
-1/x^{n},n\geq 1,x>0$. Now let us state the following auxiliary result.
\begin{lemma}
\label{th:QoSExistsWeightVec}
Suppose that \ref{as:UserPC_Continuous} and \ref{as:UserPC_LogConvex}
hold. Let $\vmat$ be irreducible. Then, $\pv\in\partial\freg$ if and only if
there exists $\ve{w}>0$ such that $\pv$ maximizes
$\ve{x}\mapsto\ve{w}^T\ve{x}$ over $\menge{F}$.
\end{lemma}
\begin{IEEEproof}
  By convexity and downward comprehensivity of $\freg$, every boundary point
  of this set is a maximal point and it maximizes
  $\ve{x}\mapsto\ve{w}^T\ve{x}$ over $\menge{F}$ for some $\ve{w}\geq 0$
  \cite[pp. 54--58]{Boy03}. Since $\vmat$ is irreducible, it follows from
  \cite[Theorem 4.3]{StISIT2006} (see also \cite[Corollary 4.3]{StISIT2006})
  that $\ve{w}>0$.
\end{IEEEproof}

Now we can easily observe the following.

\begin{observation}
\label{obs:UserPC_MaxMin_SIRBalancedUniq}
If $\vmat\geq 0$ is irreducible, then $\bar{\ve{p}}$ is unique and equal to
$\bar{\ve{p}}'$ defined by (\ref{eq:UserPC_MaxMin_SIRBalancedSpecial}).
\end{observation}
\begin{IEEEproof}
The proof is deferred to Appendix \ref{app:proofObsSIRBalancedUniq}.
\end{IEEEproof}

We complete this section by pointing out a connection between the max-min
SIR-balancing problem considered in this paper and the notion of max-min
fairness. A vector of balanced SIRs with entries $\sir_k(\ve{p}'')/\gamma_k,\
k\in\logic$ is max-min fair if any $\sir_k(\ve{p}'')/\gamma_k$ cannot be
increased without decreasing some $\sir_l(\ve{p}'')/\gamma_l,\ l\neq k$ which
is smaller than or equal to $\sir_k(\ve{p}'')/\gamma_k$; the vector $\ve{p}''$
is then a max-min fair power allocation \cite{Bertsekas92},
\cite{MoWalrand00}.  A max-min fair power allocation is therefore also max-min
SIR-balanced (provided that the feasible SIR region is downward
comprehensive); the converse is in general not true.  However, if the max-min
SIR-balancing problem has a unique solution given by
\eqref{eq:UserPC_MaxMin_SIRBalancedSpecial}, this solution is also max-min
fair and there are no other max-min fair power allocations.  These relations
can be observed in Fig. \ref{fig:UserPC_maxmin_problem} where both max-min
SIR-balanced and max-min fair points are indicated.

\section{Characterization of Max-Min SIR-balancing}
\label{sec:UserPC_MaxMin_IndCon}

In this section, we characterize $\bar{\ve{p}}\in\ppset$ defined by
(\ref{eq:UserPC_MaxMin_SIRBalancedDef}) under the assumption that $\vmat$ is
irreducible. We point out possible extensions to reducible matrices at the end
of this section. We assume that $\pset\subset\RN^K$ is a convex polytope given
by (\ref{eq:NetMod_PowerConstrSetMatrixForm}). So, throughout this section,
$\max_{n\in\nodes} g_n(\ve{p})\leq 1$ where
\begin{equation}
  \label{eq:UserPC_MaxMinGenCon_GenPConst}
  g_n(\ve{p}):=1/P_n\ve{c}^T_{n}\ve{p},\quad n\in\nodes
\end{equation}
and $\ve{c}_n\in\{0,1\}^{K}$ is a (column) vector equal to the $n$th row of
the matrix $\ma{C}$. Now, using (\ref{eq:UserPC_MaxMinGenCon_GenPConst}), the
max-min SIR power vector $\bar{\ve{p}}$ defined by
(\ref{eq:UserPC_MaxMin_SIRBalancedDef}) can be written as
\begin{align}
\label{eq:UserPC_MaxMinGenCon_PVecDef}
\bar{\ve{p}}=\underset{\ve{p}\geq
  0}{\arg\max}\min_{k\in\logic}(\sir_k(\ve{p})/\gamma_k) &&\text{s.t.}
&& \max_{n\in\nodes} g_n(\ve{p})\leq 1\,.
\end{align}
where $\gamma_k>0,k\in\logic$, is arbitrary but fixed.
\begin{lemma}
\label{lem:UserPC_MaxMinGenCon_BasicLemma}
Let $\bar{\ve{p}}$ be any power vector that solves
(\ref{eq:UserPC_MaxMinGenCon_PVecDef}). Then, the following holds
\begin{enumerate}[(i)]
\item $\max_{n\in\nodes} g_n(\bar{\ve{p}})=1$.
\item If $\vmat\geq 0$ is irreducible, then $\bar{\ve{p}}$ is unique and
\begin{equation}
\label{eq:UserPC_MaxMinGenCon_AllEqualVirreducible}
\forall_{k\in\logic}\,\gamma_k/\sir_k(\bar{\ve{p}})=\beta,\;\beta>0\,.
\end{equation}
\end{enumerate}
\end{lemma}
\begin{IEEEproof}
The proof can be found in Appendix \ref{app:proofLemmaBasicLemma}.
\end{IEEEproof}

Because $\bar{\ve{p}}$ maximizes $\min_{k\in\logic}(\sir_k(\ve{p})/\gamma_k)$
over $\pset$, it follows from
(\ref{eq:UserPC_MaxMinGenCon_AllEqualVirreducible}) that $1/\beta>0$ is the
corresponding maximum. It must be emphasized that
(\ref{eq:UserPC_MaxMinGenCon_AllEqualVirreducible}) is not true for general
nonnegative matrices $\vmat\geq 0$.  In the lemma, we require that the gain
matrix be irreducible, which is sufficient for
(\ref{eq:UserPC_MaxMinGenCon_AllEqualVirreducible}) to hold but not necessary.
The irreducibility property ensures that, regardless of the choice of $\pset$,
there is no subnetwork being completely decoupled from the rest of the
network. To be more precise, if $\vmat$ is irreducible, then the network is
entirely coupled by interference so that the type of power constraints is
irrelevant for this issue (see also the remark at the end of this
section). Unless otherwise stated, it is assumed in the remainder of this
section that $\vmat\geq 0$ is an arbitrary \emph{irreducible matrix}. Due to
$(ii)$ of Lemma \ref{lem:UserPC_MaxMinGenCon_BasicLemma}, this implies that
the max-min SIR-balanced power vector is unique.

Let us define
\begin{equation}
  \label{eq:UserPC_MaxMinGenCon_n0Def}
  \nodes_0(\ve{p}):=\bigl\{n_0\in\nodes: n_0=\underset{n\in\nodes}{\arg\max}\,g_n(\ve{p})
=1\bigr\}
\end{equation}
which includes the indices of those nodes for which the power constraints are
active under the power vector $\ve{p}$. By $(i)$ of Lemma
\ref{lem:UserPC_MaxMinGenCon_BasicLemma}, the cardinality of
$\nodes_0(\bar{\ve{p}})$ must be larger than or equal to $1$.  In what
follows, let $\beta>0$ be the constant in part $(ii)$ of the lemma. This
together with part $(i)$ implies that
\begin{align}
  \label{eq:UserPC_MaxMinSumCon_2Eqs_1}
  \beta\bar{\ve{p}}=\gmat\vmat\bar{\ve{p}}+\gmat\ve{z} &&
  g_n(\bar{\ve{p}})=1\,.
\end{align}
Putting the first equation into the second one yields a set of $K+1$ equations
that, if written in a matrix form, show that if $\bar{\ve{p}}$ solves the
max-min SIR-balancing problem, then there is a constant $\beta>0$ such
that 
\begin{equation}
\label{eq:UserPC_MaxMinGenCon_1MatEq}
\beta\,\tilde{\ve{p}}=\ma{A}^{(n)}\tilde{\ve{p}},\quad \beta>0,\tilde{\ve{p}}\in\RP^{K+1}
\end{equation}
for each $n\in\nodes_0(\bar{\ve{p}})$ where $\tilde{\ve{p}}=(\bar{\ve{p}},1)$
is the extended power vector and the nonnegative matrix
$\ma{A}^{(n)}\in\RN^{(K+1)\times(K+1)}$ is defined to be
\begin{equation}
\label{eq:UserPC_MaxMinGenCon_AMatDef}
\ma{A}^{(n)}=\begin{pmatrix}
  \gmat\vmat & \gmat\ve{z}\\
  \frac{1}{P_{n}}\ve{c}_n^T\gmat\vmat &
  \frac{1}{P_{n}}\ve{c}_n^T\gmat\ve{z}
\end{pmatrix},\quad n\in\nodes\,.
\end{equation}
Alternatively, we can write (\ref{eq:UserPC_MaxMinSumCon_2Eqs_1}) as
$\beta\bar{\ve{p}}=\gmat\vmat\bar{\ve{p}}+\gmat\ve{z}\cdot g_n(\bar{\ve{p}})$,
from which we obtain, for each $n\in\nodes_0(\bar{\ve{p}})$,
\begin{equation}
  \label{eq:UserPC_MaxMinGenCon_Representation_in_K}
  \begin{split}
    \beta\bar{\ve{p}}
    =\ma{B}^{(n)}\bar{\ve{p}},\quad \beta>0,\bar{\ve{p}}\in\RP^{K+1}
  \end{split}
\end{equation}
where $\ma{B}^{(n)}\in\RN^{K\times K}$ is defined to be (for each
$n\in\nodes$)
\begin{equation}
  \label{eq:UserPC_MaxMinGenCon_BmatrixDef}
  \ma{B}^{(n)}:=\gmat\vmat+\frac{1}{P_{n}}\gmat\ve{z}\ve{c}_n^T
  =\gmat\bigl(\vmat+\frac{1}{P_{n}}\ve{z}\ve{c}_n^T\bigr)
  =\gmat\tilde{\ma{V}}^{(n)}
\end{equation}
and $\tilde{\ma{V}}^{(n)}:=\vmat+\frac{1}{P_{n}}\ve{z}\ve{c}_n^T,n\in\nodes$.
So, given $\bar{\ve{p}}$ defined by (\ref{eq:UserPC_MaxMinGenCon_PVecDef}),
the equations (\ref{eq:UserPC_MaxMinGenCon_1MatEq}) and
(\ref{eq:UserPC_MaxMinGenCon_Representation_in_K}) hold for each
$n\in\nodes_0(\bar{\ve{p}})$. In other words, the solution of
(\ref{eq:UserPC_MaxMinGenCon_PVecDef}) in a network entirely coupled by
interference must satisfy (\ref{eq:UserPC_MaxMinGenCon_1MatEq}) and
(\ref{eq:UserPC_MaxMinGenCon_Representation_in_K}) for each node $n\in\nodes$
whose power constraints are active at the maximum. This is summarized in the
following lemma.

\begin{lemma}
  \label{lem:UserPC_MaxMinGenCon_Represenations}
  If $\vmat\geq 0$ is irreducible and $\bar{\ve{p}}$ solves the max-min
  SIR-balancing problem (\ref{eq:UserPC_MaxMinGenCon_PVecDef}), then
  $\bar{\ve{p}}$ satisfies both (\ref{eq:UserPC_MaxMinGenCon_1MatEq}) and
  (\ref{eq:UserPC_MaxMinGenCon_Representation_in_K}) for some $\beta>0$ and
  each $n\in\nodes_0(\bar{\ve{p}})$.
\end{lemma}

Note that the lemma is an immediate consequence of parts $(i)$ and $(ii)$ of
Lemma \ref{lem:UserPC_MaxMinGenCon_BasicLemma}, from which
(\ref{eq:UserPC_MaxMinGenCon_1MatEq}) and
(\ref{eq:UserPC_MaxMinGenCon_Representation_in_K}) follow for an arbitrary
$n\in\nodes_0(\bar{\ve{p}})$. Now we are in a position to prove the following
result. 

\begin{lemma}
\label{lem:UserPC_MaxMinGenCon_EigenBUnique}
Suppose that $\vmat\geq 0$ is irreducible. Then, for any constants $c_1>0$ and
$c_2>0$, the following holds.
\begin{enumerate}[(i)]
\item For each $n\in\nodes$, there is exactly one positive vector
  $\ve{p}=\ve{p}^{(n)}\in\RN^K$ with $g_n(\ve{p})=c_1$ satisfying
  $\beta^{(n)}\,\ve{p}=\ma{B}^{(n)}\ve{p}$ for some $\beta^{(n)}>0$.
  Moreover, $\beta^{(n)}$ is a simple eigenvalue of $\ma{B}^{(n)}$ and
  $\beta^{(n)}=\rho(\ma{B}^{(n)})$.
\item For each $n\in\nodes$, there is exactly one positive vector
  $\tilde{\ve{p}}=\tilde{\ve{p}}^{(n)}\in\RN^{K+1}$ with $\tilde{p}_{K+1}=c_2$
  satisfying $\beta^{(n)}\,\tilde{\ve{p}}=\ma{A}^{(n)}\tilde{\ve{p}}$ for some
  $\beta^{(n)}>0$.  Moreover, $\beta^{(n)}$ is a simple eigenvalue of
  $\ma{A}^{(n)}$ and $\beta^{(n)}=\rho(\ma{A}^{(n)})$.
\end{enumerate}
\end{lemma}
\begin{IEEEproof}
The reader can find the proof in Appendix \ref{app:proofLemmaEigenBUniq}.
\end{IEEEproof}

The lemma says that, for each $n\in\nodes$, the matrix equation
$\beta^{(n)}\,\ve{p}=\ma{B}^{(n)}\ve{p}$ with $\beta^{(n)}>0$ and
$\ve{p}\in\RN^K$ is satisfied if and only if $\ve{p}$ is a positive right
eigenvector of $\ma{B}^{(n)}$ associated with
$\beta^{(n)}=\rho(\ma{B}^{(n)})$. Furthermore, if $g_n(\ve{p})=1$, then
$\ve{p}$ is unique. Similarly, $\beta^{(n)}\,\ve{p}=\ma{A}^{(n)}\ve{p}$ with
$\beta^{(n)}>0$ and $\ve{p}\in\RN^{K+1}$ is satisfied if and only if $\ve{p}$
is a positive right eigenvector of $\ma{A}^{(n)}$ associated with
$\beta^{(n)}=\rho(\ma{A}^{(n)})$ and there is exactly one such an eigenvector
whose last entry is equal to one.  Furthermore, for each $n\in\nodes$,
$\beta^{(n)}=\rho(\ma{A}^{(n)})=\rho(\ma{B}^{(n)})$. This is because if
$\rho(\ma{B}^{(n)})\ve{p}=\ma{B}^{(n)}\ve{p}$ holds for some $n\in\nodes$,
then we must have
$\rho(\ma{B}^{(n)})\tilde{\ve{p}}=\ma{A}^{(n)}\tilde{\ve{p}}$ where
$\tilde{\ve{p}}=(\ve{p},1)\in\RP^K$. Thus, by Lemma
\ref{lem:UserPC_MaxMinGenCon_EigenBUnique}, we must have
\begin{equation}
  \label{eq:UserPC_MaxMinGenCon_rhoA=rhoB}
  \rho(\ma{A}^{(n)})=\rho(\ma{B}^{(n)}),\quad n\in\nodes\,.
\end{equation}
Note that a solution to the max-min SIR-balancing problem is not necessarily
obtained for each $n\in\nodes$ since, in the optimum, some power constraints
may be inactive. Indeed, in general, the set
$\nodes_0^c(\bar{\ve{p}})=\nodes\setminus\nodes_0(\bar{\ve{p}})$ is not an
empty set where $\bar{\ve{p}}$ defined by
(\ref{eq:UserPC_MaxMinGenCon_PVecDef}) is unique due to $(ii)$ of Lemma
\ref{lem:UserPC_MaxMinGenCon_BasicLemma}.

Now we combine Lemmas \ref{lem:UserPC_MaxMinGenCon_Represenations} and
\ref{lem:UserPC_MaxMinGenCon_EigenBUnique} to obtain the following.
\begin{theorem}
\label{th:UserPC_MaxMinGenCon_MainTheorem}
Let $\beta$ be given by
(\ref{eq:UserPC_MaxMinGenCon_AllEqualVirreducible}). If $\vmat\geq 0$ is
irreducible, then the following statements are equivalent.
\begin{enumerate}[(i)]
\item $\bar{\ve{p}}\in\ppset$ solves the max-min SIR-balancing problem
  (\ref{eq:UserPC_MaxMinGenCon_PVecDef}).
\item For each $n\in\nodes_0(\bar{\ve{p}})$, $\bar{\ve{p}}$ is a unique
  positive right eigenvector of $\ma{B}^{(n)}$ associated with
  $\beta=\rho(\ma{B}^{(n)})>0$ such that
  $g_n(\bar{\ve{p}})=1$.
\item For each $n\in\nodes_0(\bar{\ve{p}})$, $\tilde{\ve{p}}$ is a unique
  positive right eigenvector of $\ma{A}^{(n)}$ associated with
  $\beta=\rho(\ma{A}^{(n)})>0$ such that $\tilde{p}_{K+1}=1$.
\end{enumerate}
\end{theorem}
\begin{IEEEproof}
The proof can be found in Appendix \ref{app:proofMainTheorem}.
\end{IEEEproof}

Theorem \ref{th:UserPC_MaxMinGenCon_MainTheorem} implies that if $\vmat$ is
irreducible, then $\bar{\ve{p}}>0$ is the (positive) right eigenvector of
$\ma{B}^{(n)}$ associated with $\rho(\ma{B}^{(n)})\in\sigma(\ma{B}^{(n)})$ for
each $n\in\nodes_0(\bar{\ve{p}})$. Alternatively, $\bar{\ve{p}}$ can be
obtained from $\tilde{\ve{p}}=(\bar{\ve{p}},1)$, which is the positive right
eigenvector of $\ma{A}^{(n)}$ associated with $\rho(\ma{A}^{(n)})$ for each
$n\in\nodes_0(\bar{\ve{p}})$. The problem is, however, that
$\nodes_0(\bar{\ve{p}})$ is not known as this set is determined by the
solution to the max-min SIR-balancing problem, and hence its determination is
itself a part of the problem. As the SIR targets are feasible if and only if
they are met under $\bar{\ve{p}}$, the following characterization of the set
$\nodes_0(\bar{\ve{p}})$ immediately follows from \cite{MahdaviISIT07} and
(\ref{eq:UserPC_MaxMinGenCon_rhoA=rhoB}).
\begin{theorem}[\cite{MahdaviISIT07}]
  \label{th:UserPC_MaxMinGenCon_maxRho<1}
  Suppose that $\vmat\geq 0$ is irreducible. Then,
  \begin{equation}
    \label{eq:UserPC_MaxMinGenCon_N0withMaxSpecRadii}
    \begin{split}
    \nodes_0(\bar{\ve{p}})
    &=\bigl\{n_0\in\nodes:n_0={\arg\max}_{n\in\nodes}\,\rho(\ma{B}^{(n)})\bigr\}
    \end{split}
  \end{equation}
  Moreover, the feasible SIR region $\menge{F}_{\ve{\gamma}}$ is characterized as
    \begin{align}
      \label{eq:UserPC_MaxMinGenCon_maxRho<1}
      \menge{F}_{\ve{\gamma}}=\bigl\{\ve{\gamma}\in\RP^K:
      \ &\underset{n\in\nodes}{\max}\,\rho(\ma{B}^{(n)})
      =\underset{n\in\nodes}{\max}\,\rho(\ma{A}^{(n)})
      \leq 1,\nonumber\\
      &\gmat=\diag(\gamma_1,\dotsc,\gamma_K)\bigr\}.
    \end{align}
\end{theorem}
The characterization of the feasible SIR region in \eqref{eq:UserPC_MaxMinGenCon_maxRho<1}
can be deduced directly from \cite{MahdaviISIT07} as the authors show that 
$\underset{k\in\logic}{\min}(\sir_k(\ve{p})/\gamma_k) \leq \underset{n\in\nodes}{\min}(1 / \rho(\ma{B}^{n}))$,
without characterizing, however, the corresponding power allocation vector $\ve{p}$.

Theorems \ref{th:UserPC_MaxMinGenCon_MainTheorem} and
\ref{th:UserPC_MaxMinGenCon_maxRho<1} directly lead to the following procedure
for computing the max-min SIR power vector $\bar{\ve{p}}$ given by
(\ref{eq:UserPC_MaxMin_SIRBalancedDef}).
\begin{algorithm}
\label{alg:UserPC_MaxMinGenPCConNovel}
\begin{algorithmic}[1]
  \REQUIRE $\gmat=\diag(\gamma_1,\dotsc,\gamma_K)$.
\ENSURE $\bar{\ve{p}}\in\ppset\subset\pset$
\STATE Find an arbitrary index $n_0\in\nodes$ such that
  $n_0=\arg\max_{n\in\nodes}\rho(\ma{B}^{(n)})$ where $\ma{B}^{(n)}$ is given
  by (\ref{eq:UserPC_MaxMinGenCon_BmatrixDef}). 
\STATE Let $\bar{\ve{p}}$  be given by 
  $\rho(\ma{B}^{(n_0)})\bar{\ve{p}}=\ma{B}^{(n_0)}\bar{\ve{p}}$ and normalized such that
  $\ve{c}_{n_0}^T\bar{\ve{p}}=P_{n_0}$. 
\end{algorithmic}
\end{algorithm}

\begin{remark}[Remark on reducible matrices]
  We point out that all the statements in this paper hold if $\ma{B}^{(n)}$ is
  irreducible for each $n\in\nodes$, which may be satisfied even if $\vmat$ is
  reducible. This is for instance true in the case of a sum power constraint
  ($\ma{C}=(1,\dotsc,1)$) where $\ma{B}=\ma{B}^{(1)},\nodes=\{1\},$ is a
  positive matrix. Moreover, the statement of Theorem
  \ref{th:UserPC_MaxMinGenCon_MainTheorem} can be shown to be true even if
  $\ma{B}^{(n)}$ is irreducible only for $n\in\nodes_0(\bar{\ve{p}})$.
\end{remark}



\section{Applications}
\label{sec:applications}

In this section, we discuss two other applications of the results. In doing so, $\vmat$
 is assumed to be irreducible. 
 Under this assumption, the feasible QoS region given by \eqref{eq:UserPC_MaxMin_FeasibleQoSRegion}
 can be shown to be strictly convex \cite[Corrolary 4.3]{StISIT2006}. 



\subsection{Computation via utility-based power control}
\label{sec:appl_utilityPC}

In this section, we show that $\bar{\ve{p}}$ can be obtained by maximizing the
following aggregate utility function
\begin{equation}
  \label{eq:RateCon_SIRbalancing_ObjectiveDef}
  F(\ve{p},\ve{w}):=
  \sum\nolimits_{k\in\logic} w_k\phi\bigl(\sir_k(\ve{p})/\gamma_k\bigr) 
\end{equation}
provided that the weight vector $\ve{w}=(w_1,\dotsc,w_K)>0$ is chosen
appropriately and $\phi:\RP\to\qset$ satisfies \ref{as:UserPC_Continuous} and
\ref{as:UserPC_LogConvex}. To this end, define
\begin{equation}
\label{eq:UserPC_OptUtilityPV}
\ve{p}^\ast(\ve{w}):=\arg\max\nolimits_{\ve{p}\in\ppset}\,F(\ve{p},\ve{w})\,.
\end{equation}
Although $\ppset$ is not compact, it can be shown \cite{StBookSpringer08} that
the maximum exists if \ref{as:UserPC_Continuous} and \ref{as:UserPC_LogConvex}
are fulfilled. Furthermore, it is obvious that in the maximum, at least one
power constraint is active, that is, $\ma{C}\ve{p}^\ast\leq\hat{\ve{p}}$ holds
at least with one equality. Thus, we have
$\pv^\ast(\ve{w})=(\pve_1^\ast(\ve{w}),\dotsc,\pve_K^\ast(\ve{w}))\in\partial\freg$
for
\begin{equation}
\label{eq:OptimalQoSUtility}
\pve^\ast_k(\ve{w})=\phi\bigl(\sir_k(\ve{p}^\ast(\ve{w}))/\gamma_k\bigr)\,. 
\end{equation}
In words, $\ve{p}^\ast=\ve{p}^\ast(\ve{w})$ corresponds to a boundary point of
$\freg$ defined by (\ref{eq:UserPC_MaxMin_FeasibleQoSRegion}).  Different
boundary points can be achieved by choosing different weight vectors in
(\ref{eq:RateCon_SIRbalancing_ObjectiveDef}). In particular, Lemma
\ref{th:QoSExistsWeightVec} implies that $\ve{q}\in\partial\freg$ if and only
if $\ve{q}=\ve{q}^\ast(\ve{w})$ for some $\ve{w}>0$. For the analysis in this
section, it is important to recall from Observation \ref{obs:F_Properties}
that there is a bijective map from $\freg$ onto $\pset$ and this map can be
shown to be \cite[Section 5.3]{StBookSpringer08}
\begin{equation}
\label{eq:bijective_map}
\ve{p}(\pv)=(\ma{I}-\ma{G}(\pv)\gmat\vmat)^{-1}\ma{G}(\pv)\gmat\ve{z},\;\pv\in\freg
\end{equation}
where $\ma{G}(\pv):=\diag(g(\pve_1),\dotsc,g(\pve_K))$ with $g(x)$ defined by 
\ref{as:UserPC_LogConvex} and $\rho(\ma{G}(\pv)\gmat\vmat)<1$, which ensures 
the existence of $\ve{p}(\pv)$
and is satisfied for every $\pv\in\freg$ \cite[Section
5.3]{StBookSpringer08}. Note that $g$ is strictly increasing and $g(x)>0$ for
all $x\in\qset$, which follows from \ref{as:UserPC_Continuous}
and \ref{as:UserPC_LogConvex}. 
\begin{lemma}
  \label{lem:UserPC_MaxMinrho=1Boundary}
   $\pv\in\partial\freg$ if and only if
  $\max_{n\in\nodes}\perron_n(\pv)=1$ where
  $\perron_n(\ve{q}):=\rho(\ma{G}(\pv)\ma{B}^{(n)})>\rho(\ma{G}(\pv)\gmat\vmat)>0$.
\end{lemma}
\begin{IEEEproof}
The proof is deferred to Appendix \ref{app:proofLemmarho=1Boundary}.
\end{IEEEproof}
Now we are in a position to prove the following.
\begin{theorem}
\label{th:SpecificationWeight}
Suppose that \ref{as:UserPC_Continuous} and \ref{as:UserPC_LogConvex} hold.
Let $\pv\in\partial\freg$ and
$\ve{u}(\pv)=(g'(\pve_1)/g(\pve_1),\dotsc,g'(\pve_K)/g(\pve_K))>0$. Then, we
have $\pv=\pv^\ast(\ve{w})$ given by (\ref{eq:OptimalQoSUtility}) if
\begin{equation}
\label{eq:SpecificationWeight}
\ve{w} = c\cdot\ve{u}(\pv)\circ\ve{y}\circ\ve{x},\quad c>0
\end{equation}
where $\ve{y}$ and $\ve{x}$ are positive left and right eigenvectors of
$\ma{G}(\pv)\ma{B}^{(n_0)}$, respectively, associated with
$\perron_{n_0}(\pv)$ for any $n_0=\arg\max_{n\in\nodes}\perron_n(\pv)$.
\end{theorem}
\begin{IEEEproof}
The reader can find the proof in Appendix \ref{app:proofTheoremSpecificationWeight}.
\end{IEEEproof}

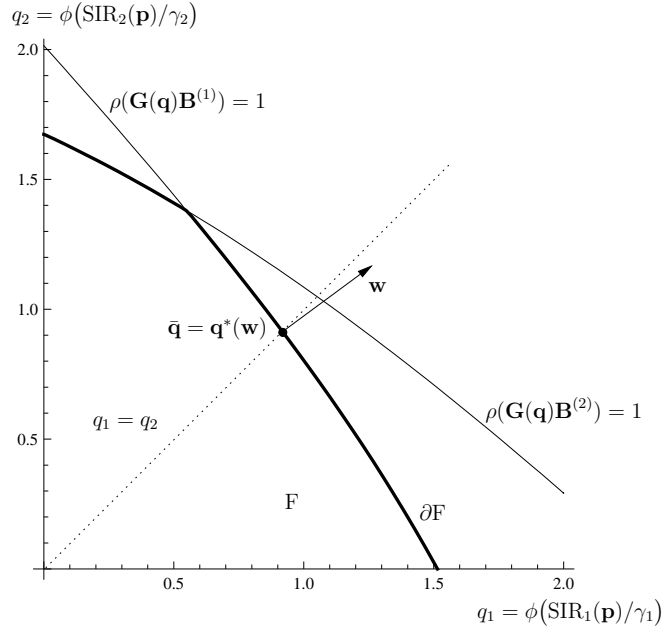
\begin{figure}[htbp]
\centering
  \scalebox{0.7}{\input{feasibleQoS.pstex_t}}
\caption{An illustration of Corollary \ref{cor:SpecificationWeightMaxMin}. The
  figure shows an example of the feasible QoS region defined by
  (\ref{eq:UserPC_MaxMin_FeasibleQoSRegion}) with $2$ users subject to
  individual power constraints ($\ma{C}=\ma{I}$). The point
  $\bar{\pv}=\pv^\ast(\ve{w})$ with $\ve{w}=c\ve{y}\circ\ve{x},c>0$,
  corresponds to the unique max-min SIR-balanced power allocation. The weight
  vector $\ve{w}$ is normal to a hyperplane which supports the feasible QoS
  region at $\bar{\pv}\in\partial\freg$. }
\label{fig:obrazek}
\end{figure}

Now we can establish a connection between
(\ref{eq:UserPC_MaxMin_SIRBalancedDef}) and
(\ref{eq:UserPC_OptUtilityPV}). The connection is illustrated by Figure
\ref{fig:obrazek}. 

\begin{corollary}
\label{cor:SpecificationWeightMaxMin}
Let $\ve{y}$ and $\ve{x}$ be positive left and right eigenvectors of
$\ma{B}^{(n_0)}$ associated with $\rho(\ma{B}^{(n_0)})$ for any
$n_0\in\nodes_0(\bar{\ve{p}})$. If $\ve{w}=c\,\ve{y}\circ\ve{x},c>0$, then
$\bar{\ve{p}}=\ve{p}^\ast(\ve{w})$.
\end{corollary}
\begin{IEEEproof}
The proof can be found in Appendix \ref{app:proofCorSpecifWeight}.
\end{IEEEproof}

\subsection{Saddle point characterization of the Perron roots}
\label{sec:appl_saddlePC}

Finally, we point out that Theorem \ref{th:UserPC_MaxMinGenCon_MainTheorem}
gives rise to a saddle point characterization of
$\rho(\ma{B}^{(n)}),n\in\nodes_0(\bar{\ve{p}})$. 
Let $\simplex:=\{\ve{x}\in\RN^K:\|\ve{x}\|_1=1\}$ and
$\simplexp=\simplex\cap\RP^K$. We define
$G:\simplexp\times\ppset\to\R$ as
\begin{equation*}
G(\ve{w},\ve{p}):=\sum\nolimits_{k\in\logic}
w_k\psi\bigl(\gamma_k/\sir_k(\ve{p})\bigr) 
\end{equation*}
where $\psi(x):=-\phi(1/x)$. 
A key ingredient in the proof of the saddle point characterization is the 
following theorem, which can be deduced from \cite[Sections 1.2.4--5]{StBookSpringer08}:
\begin{theorem}
\label{th:upperBoundonPsi(rho)}
Assume that \ref{as:InterferenceLinear}--\ref{as:UserPC_LogConvex} hold 
and $\vmat$ is irreducible. Let
$\ma{B}=\ma{B}^{(n)}$ for any $n\in\nodes$, and let
$\ve{w}=\ve{y}\circ\ve{x}\in\simplexp$ where $\ve{y}$ and $\ve{x}$ are
positive left and right eigenvectors of $\ma{B}$ associated with $\rho(\ma{B})$. 
Then, for all $\ve{p}>0$,
\begin{equation}
\label{eq:upperBoundonPsi(rho)}
\psi(\rho(\ma{B}))\leq\sum\nolimits_{k\in\logic}w_k\psi\Bigl(\frac{(\ma{B}\ve{p})_k}{p_k}\Bigr)\,.
\end{equation}
Equality holds if and only if $\ve{p}=\ve{x}>0$. 
\end{theorem}


Now we are in a position to present the saddle point characterization
of the Perron root of $\ma{B}^{(n)},n\in\nodes_0(\bar{\ve{p}})$, 
similar to the one for the noiseless case 
which can be found in \cite{StBookSpringer08}. 

\begin{theorem}
\label{th:saddlepoint}
Suppose that \ref{as:UserPC_Continuous} and
\ref{as:UserPC_LogConvex} hold, and $\vmat$ is irreducible. Then,
\begin{equation*}
  \psi(\rho(\ma{B}^{(n_0)}))
  =\sup_{\ve{w}\in\simplexp}\inf_{\ve{p}\in\ppset}G(\ve{w},\ve{p})
  =\inf_{\ve{p}\in\ppset}\sup_{\ve{w}\in\simplexp}G(\ve{w},\ve{p})  
\end{equation*}
where $n_0=\arg\max_{n\in\nodes}\rho(\ma{B}^{(n)})$.
Moreover, a point $(\ve{w}^\ast,\ve{p}^\ast)$ is the saddle point of 
$G(\ve{w},\ve{p})$ if and only if $\ve{p}^\ast=\bar{\ve{p}}$ and $\ve{w}\in\menge{W}$,
where
\begin{equation*}
  \label{eq:ApplutilityMax_WsetDef}  
\menge{W}=\Bigl\{\ve{w}\in\simplexp:\ve{w}=\sum_{n\in\nodes_0}c_n\ve{w}^{(n)},
\sum_{n\in\nodes_0}c_n=1,
c_n\geq 0\Bigr\}
\end{equation*}
and $\ve{w}^{(n)}:=\ve{y}^{(n)}\circ\ve{x}^{(n)}\in\simplexp$ with
$\rho(\ma{B}^{(n)})\ve{x}^{(n)}=\ma{B}^{(n)}\ve{x}^{(n)}$,
$\rho(\ma{B}^{(n)})\ve{y}^{(n)}=(\ma{B}^{(n)})^T\ve{y}^{(n)}$, 
$(\ve{y}^{(n)})^T\ve{x}^{(n)}=1$.
In words, at the saddle point the power vector is equal to the max-min SIR-balanced 
power allocation, whereas the weight vector is any linear combination of 
the vectors $\ve{w}^{(n)}$ for $n\in\nodes_0(\bar{\ve{p}})$.
\end{theorem}

With Theorems \ref{th:UserPC_MaxMinGenCon_MainTheorem} and
\ref{th:upperBoundonPsi(rho)} 
in hand, the proof is similar to that in \cite[Section 1.2.4]{StBookSpringer08}.
The existence of a saddle point is ensured by irreducibility of the gain
matrix since then positive left and right eigenvectors exist. The uniqueness
follows from the irreducibility property and the normalizations. 

The reason why the theorem is of interest is that it provides a basis for the
design of alternative power control algorithms for saddle point problems that
converge to $\bar{\ve{p}}$ and may be amenable to distributed
implementation. Basically, the idea of the algorithm is redolent of that of 
primal-dual algorithms that converge to a saddle point of the associated Lagrange function
\cite{Boy03}. 
Development of new algorithms is currently a subject of our ongoing work; the main idea,
however, consists in minimizing the function $G(\ve{w},\ve{p})$ with respect to $\ve{p}$,
and \emph{simultaneously} maximizing $G(\ve{w},\ve{p})$ with respect to $\ve{w}$. 
The straight-forward approach employs the gradient projection method, where in order 
for the objective function to be convex in the power variable the substitution
$\ve{s}=\log(\ve{p})$ is used. Each iteration encompasses the calculation of 
the gradient of $G(\ve{w},e^{\ve{s}})$, and an update of the vectors $\ve{w}$ and
$\ve{s}$ in the direction and against the direction of the gradient,
respectively. This process requires a suitable step size to be chosen. The iteration 
is concluded with a projection of the updated values of $\ve{w}$ and
$\ve{s}$ onto the corresponding sets of valid values.

The minimization of $G(\ve{w},\ve{p})$ in the power domain using the gradient
projection method corresponds to employing the power control algorithm
presented in \cite[Section 6.5]{StBookSpringer08}. In particular, the gradient
can be computed in a distributed way, and the projection onto the feasible set
is also distributedly implementable in many cases of interest. As for the optimization
in the weights domain, the gradient can be computed independently by each node,
but performing the projection requires in general centralized operation.

\section{Conclusions}
\label{sec:concl}

In this paper, we have characterized the max-min SIR-balanced power allocation under
general power constraints. This characterization is an extension of the results
known previously for the noiseless case, in which the power constraints play no role.
We have also established a connection between the max-min SIR-balancing power allocation
problem and the utility-based power allocation problem for the considered case, 
and as an application of our results we have discussed two classes of power allocation 
algorithms based on those two approaches. Finally, we have presented a saddle-point 
characterization of the Perron roots of the extended gain matrices which may constitute 
a basis for developing distributed power control algorithms.

\section{Acknowledgments}
\label{sec:acknowl}

%
We thank our colleagues Dr. Chee Wei Tan, Prof. Mung Chiang and Prof.
R. Srikant for bringing to our attention their Infocom'09 paper
\cite{TanChiangSrikant09}, after the publication of our technical report
\cite{StKaliszISIT09arxiv} on which the present paper is fully based. In their
paper \cite{TanChiangSrikant09}, they have independently obtained some related
results.  We also refer the reader to \cite{Chiang08PowerControl} for
additional views on the issues.

Finally, as explained in \cite{StBookSpringer08}, Theorem
\ref{th:upperBoundonPsi(rho)} of this paper is substantially related to
results of the seminal paper of Friedland \& Karlin \cite{FriedlandKarlin75}
to which also the reader is referred.

\section{Appendix}
\label{sec:Appendix}

\subsection{Proof of Observation \ref{obs:UserPC_MaxMin_SIRBalancedUniq}}
\label{app:proofObsSIRBalancedUniq}

  Let \ref{as:UserPC_LogConvex} be satisfied, and let $\bar{\ve{p}}\in\ppset$
  be any solution to (\ref{eq:UserPC_MaxMin_SIRBalancedDef}) or, equivalently,
  (\ref{eq:UserPC_MaxMin_SIRBalancedInFQR}). Let
  $\bar{\pve}_k=\phi(\sir_k(\bar{\ve{p}})/\gamma_k),k\in\logic$ . By
  Observation \ref{obs:F_Properties}, $\menge{F}$ is a convex downward
  comprehensive set and, by \ref{as:UserPC_Continuous},
  (\ref{eq:UserPC_MaxMin_SIRBalancedInFQR}) and
  (\ref{eq:UserPC_MaxMin_FeasibleQoSRegion}),
  $\bar{\pv}=(\bar{\pve}_1,\dotsc,\bar{\pve}_K)\in\partial\menge{F}$ is its
  boundary point since at least one power constraint is active in the optimum
  (see Lemma \ref{lem:UserPC_MaxMinGenCon_BasicLemma}). Thus, by
  irreducibility of $\vmat$ and Lemma \ref{th:QoSExistsWeightVec}, there
  exists $\ve{w}>0$ such that $\ve{w}^T(\bar{\pv}-\ve{u})\geq 0$ for all
  $\ve{u}\in\menge{F}$. Due to positivity of $\ve{w}$, this implies that for
  any $\ve{u}\in\menge{F},\ve{u}\neq\bar{\pv}$, there is at least one index
  $i=i(\bar{\pv}, \ve{u})\in\logic$ such that $\bar{\pve}_{i}>u_{i}$. In particular,
  for any $\ve{u}\in\freg,\ve{u}\neq\bar{\pv}'$ there is $i=i(\bar{\pv}',\ve{u})\in\logic$ 
  such that $\bar{\pve}_{i}'>u_i$ where
  $\bar{\pve}'_k=\phi(\sir_k(\bar{\ve{p}}')/\gamma_k),k\in\logic$. On the
  other hand, however, we have $\bar{\pv}'\leq\bar{\ve{\pv}}$. This is simply
  because
  $\bar{\pve}'_1=\dotsb=\bar{\pve}'_K=\min_{k\in\logic}\phi(\sir_k(\bar{\ve{p}}))$. Combining
  both inequalities shows that $\bar{\pv}=\bar{\pv}'$, and hence, by
  bijectivity, we obtain $\bar{\ve{p}}=\bar{\ve{p}}'$, which is unique by
  \cite[Theorem A.51]{StBookSpringer08}.

\subsection{Proof of Lemma \ref{lem:UserPC_MaxMinGenCon_BasicLemma}}
\label{app:proofLemmaBasicLemma}

Part (i) should be obvious since if we had $g_n(\bar{\ve{p}})<1$ for all
$n\in\nodes$, then it would be possible to increase
$\min_{k\in\logic}\sir_k(\bar{\ve{p}})/\gamma_k$ by allocating the power
vector $c\bar{\ve{p}}\in\ppset$ with
$c=1/\max_{n\in\nodes}g_n(\bar{\ve{p}})>1$. In order to show part $(ii)$, note
that if \ref{as:UserPC_Continuous} and \ref{as:UserPC_LogConvex} hold, then,
by Observation \ref{obs:F_Properties}, $\fset$ is a convex downward
comprehensive set. Moreover, $\bar{\ve{p}}\in\ppset$ given by
(\ref{eq:UserPC_MaxMin_SIRBalancedInFQR}) corresponds to a point
$\bar{\pv}\in\partial\fset$, with
$\bar{\pve}_k=\phi(\sir_k(\bar{\ve{p}})/\gamma_k),k\in\logic$. Thus, by
irreducibility of $\vmat$, it follows from Lemma \ref{th:QoSExistsWeightVec}
that $\bar{\pv}$ is a maximal point of $\fset$, and hence $\bar{\pv}\leq\pv$
for any $\pv\in\fset$ implies that $\pv=\bar{\pv}$ \cite{Boy03}.  That is,
there is no vector in $\fset$ that is larger in all components than
$\bar{\pv}$. On the other hand, by the discussion in
Sect. \ref{sec:UserPC_MaxMin_Uniqueness}, $\bar{\pv}$ is a point where the
hyperplane in the direction of the vector $(1/K,\dotsc,1/K)$ intersects the
boundary of $\fset$. As a result, $\bar{\pve}_1=\dotsb=\bar{\pve}_K$, which
together with the maximality property and strict monotonicity of $\phi$, shows
that $\sir_k(\bar{\ve{p}})/\gamma_k=\beta$ for each $k\in\logic$ where $\beta$
is positive due to $(i)$. If $\vmat$ is irreducible, the uniqueness of
$\bar{\ve{p}}$ follows from Observation
\ref{obs:UserPC_MaxMin_SIRBalancedUniq}.

\subsection{Proof of Lemma \ref{lem:UserPC_MaxMinGenCon_EigenBUnique}}
\label{app:proofLemmaEigenBUniq}

  Let $n\in\nodes$ be arbitrary. First we prove part $(i)$. Since
  $1/P_n\ve{z}\ve{c}_n^T\geq 0$ and $\vmat$ is irreducible, we can conclude
  from (\ref{eq:UserPC_MaxMinGenCon_BmatrixDef}) that $\ma{B}^{(n)}\geq 0$ is
  irreducible as well.  Thus, by the Perron-Frobenius theorem for irreducible
  matrices \cite{Me00,Ho85}, there exists a positive vector $\ve{p}$ which is 
  an eigenvector of $\ma{B}^{(n)}$ associated with $\rho(\ma{B}^{(n)})$,
  and there are no nonnegative eigenvectors of $\ma{B}^{(n)}$
  associated with $\rho(\ma{B}^{(n)})$ other than $\ve{p}$ and its positive
  multiples. Among all the positive eigenvectors, there is exactly one
  eigenvector $\ve{p}>0$ such that $g_n(\ve{p})=c_1$. This proves part
  $(i)$. In order to prove $(ii)$, note that if $\ma{A}^{(n)}$ was
  irreducible, then we could invoke the Perron-Frobenius theorem and proceed
  essentially as in part $(i)$ to conclude $(ii)$ (with the uniqueness
  property resulting from the normalization of the eigenvector so that its
  last component is equal to $c_2>0$). In order to show that $\ma{A}^{(n)}$ is
  irreducible, let $\graph(\ma{A}^{(n)})$ be the associated directed graph of
  $\{1,\dotsc,K+1\}$ nodes \cite{Me00}. Since $\gmat\vmat$ is irreducible, it
  follows that the subgraph $\graph(\gmat\vmat)$ is strongly connected
  \cite{Me00}. Furthermore, as the vector $\gmat\ve{z}$ is positive, we can
  conclude from (\ref{eq:UserPC_MaxMinGenCon_AMatDef}) that there is a
  directed edge leading from node $K+1$ to each node $n<K+1$ belonging to the
  subgraph $\graph(\gmat\vmat)$. Finally, note that as $\gmat\vmat$ is
  irreducible, each row of $\gmat\vmat$ has at least one positive
  entry. Hence, the vector $1/P_n\ve{c}_n^T\gmat\vmat$ has at least one
  positive entry as well, from which and
  (\ref{eq:UserPC_MaxMinGenCon_AMatDef}) it follows that there is a directed
  edge leading from a node belonging to $\graph(\gmat\vmat)$ to node
  $K+1$. So, $\graph(\ma{A}^{(n)})$ is strongly connected, and thus
  $\ma{A}^{(n)}$ is irreducible.

\subsection{Proof of Theorem \ref{th:UserPC_MaxMinGenCon_MainTheorem}}
\label{app:proofMainTheorem}

  $(i)\to(ii)$: By Lemma \ref{lem:UserPC_MaxMinGenCon_Represenations},
  $\bar{\ve{p}}\in\ppset$ satisfies
  (\ref{eq:UserPC_MaxMinGenCon_Representation_in_K}) for some $\beta>0$. Thus,
  by Lemma \ref{lem:UserPC_MaxMinGenCon_EigenBUnique}, part $(i)$ implies part
  $(ii)$. $(ii)\to(iii)$: Given any $n\in\nodes_0(\bar{\ve{p}})$, it follows
  from (\ref{eq:UserPC_MaxMinGenCon_Representation_in_K}) that
  $\rho(\ma{B}^{(n)})\bar{\ve{p}}=\ma{B}^{(n)}\bar{\ve{p}}$ with
  $g_n(\bar{\ve{p}})=1$ is equivalent to
  $\rho(\ma{B}^{(n)})\bar{\ve{p}}=\gmat\vmat\bar{\ve{p}}+\gmat\ve{z}$, which
  in turn can be rewritten to give (\ref{eq:UserPC_MaxMinGenCon_1MatEq}) with
  $\beta=\rho(\ma{B}^{(n)})$ and $\tilde{\ve{p}}=(\bar{\ve{p}},1)$. Since
  $\bar{\ve{p}}$ is positive, so is also $\tilde{\ve{p}}$. Thus,
  $\tilde{\ve{p}}$ with $\tilde{p}_{K+1}=1$ is a positive right eigenvector of
  $\ma{A}^{(n)}$ and the associated eigenvalue is equal to
  $\rho(\ma{B}^{(n)})>0$. So, considering part $(iii)$ of Lemma
  \ref{lem:UserPC_MaxMinGenCon_EigenBUnique}, we can conclude that $(iii)$
  follows from $(ii)$. $(iii)\to(i)$: By Lemma
  \ref{lem:UserPC_MaxMinGenCon_EigenBUnique}, for each
  $n\in\nodes_0(\bar{\ve{p}})$, there exists exactly one positive vector
  $\tilde{\ve{p}}$ with $\tilde{p}_{K+1}=1$ such that
  (\ref{eq:UserPC_MaxMinGenCon_1MatEq}) is satisfied. Furthermore,
  $\beta=\rho(\ma{A}^{(n)}),n\in\nodes_0(\bar{\ve{p}})$ is a simple eigenvalue
  of $\ma{A}^{(n)}$. Now considering Lemma
  \ref{lem:UserPC_MaxMinGenCon_Represenations} proves the last missing
  implication.

\subsection{Proof of Lemma \ref{lem:UserPC_MaxMinrho=1Boundary}}
\label{app:proofLemmarho=1Boundary}

  By (\ref{eq:UserPC_MaxMin_FeasibleQoSRegion}) with
  \ref{as:UserPC_Continuous}, we have $\pv\in\freg$ if and only if there
  is $\ve{p}\in\pset$ such that $\phi(\sir_k(\ve{p})/\gamma_k)\geq\pve_k$ for
  each $k\in\logic$. Thus, $\pv\in\freg$ if and only if
  $1/\lambda:=\max_{\ve{p}\in\pset}\min_{k\in\logic}(\sir_k(\ve{p})/\gamma_kg(\pve_k))\geq
  1$ where the maximum always exists. Comparing the left hand side of the
  inequality above with (\ref{eq:UserPC_MaxMin_SIRBalancedDef}) shows that the
  only difference to the original problem formulation is that $\gamma_k$ is
  substituted by $\gamma_kg(\pve_k)$ or, equivalently, $\gmat$ by
  $\ma{G}(\pv)\gmat$, which is positive definite as well. Thus, by
  (\ref{eq:UserPC_MaxMinGenCon_AllEqualVirreducible}), Theorem
  \ref{th:UserPC_MaxMinGenCon_MainTheorem} and Theorem
  \ref{th:UserPC_MaxMinGenCon_maxRho<1}, we have $\pv\in\freg$ if and only if
  $\lambda=\max_{n\in\nodes}\perron_n(\pv)\leq 1$. Moreover, $\ve{p}(\pv)$
  given by (\ref{eq:bijective_map}) is the \emph{unique} power vector such
  that $\pve_k=\phi(\sir_k(\ve{p}(\pv))/\gamma_k)$ for each
  $k\in\logic$. Since the Neumann series converges for any $\pv\in\freg$, we
  have
  $\ve{p}(\pv)=\sum_{l=0}^\infty(\ma{G}(\pv)\gmat\vmat)^l\ma{G}(\pv)\gmat\ve{z}$. Now
  as $\ma{G}(\pv)\gmat\ve{z}$ is positive and $\ma{G}(\pv)\gmat\vmat$ is
  irreducible, we can conclude from \cite[Lemma A.28]{StBookSpringer08} and
  \ref{as:UserPC_Continuous} that each entry of $\ve{p}(\pv)$ is
  \emph{strictly increasing} in each entry of $\pv$. Thus, as $\freg$ is
  downward comprehensive and $\pv\notin\partial\freg$ holds if and only if all
  power constraints are inactive, for every $\pv\in\interior(\freg)$, there is
  $\tilde{\pv}\in\partial\freg$ such that $\tilde{\pv}=\pv+\ve{u}$ for some
  $\ve{u}>0$. By irreducibility of $\ma{B}^{(n)}$, this implies that
  $\perron_n(\pv)<\perron_n(\pv+\ve{u})=\perron_n(\tilde{\pv})\leq 1$ for each
  $n\in\nodes$ So, if $\max_{n\in\nodes}\perron_n(\pv)=1$, then
  $\pv\in\partial\freg$. Conversely, if $\pv\in\partial\freg$, we must have
  $\max_{n\in\nodes}\perron_n(\pv)=1$ since otherwise there would exist
  $\tilde{\pv}\notin\freg$ such that
  $\max_{n\in\nodes}\perron_n(\tilde{\pv})=1$, which would contradict Theorem
  \ref{th:UserPC_MaxMinGenCon_maxRho<1}. This completes the proof.

\subsection{Proof of Theorem \ref{th:SpecificationWeight}}
\label{app:proofTheoremSpecificationWeight}

  Let $\tilde{\pv}\in\partial\freg$ and
  $n_0=\arg\max_{n\in\nodes}\perron_n(\pv)$ be arbitrary and note that $\freg$
  is a convex set. So, by Lemma \ref{th:QoSExistsWeightVec}, there is
  $\ve{w}>0$ such that $\tilde{\pv}$ maximizes $\pv\mapsto\ve{w}^T\pv$ over
  $\freg$. Lemma \ref{lem:UserPC_MaxMinrho=1Boundary} implies that this convex
  problem can be stated as $\tilde{\pv}=\arg\max_{\pv}\ve{w}^T\pv$ subject to
  $\perron_{n_0}(\pv)=1,\pv\in\qset^K$. Due to \ref{as:UserPC_Continuous}, the
  spectral radius is continuously differentiable on $\qset^K$. Thus, the
  Karush-Kuhn-Tucker conditions \cite{Boy03}, which are necessary and
  sufficient for optimality (due to the convexity property), imply that
  $\ve{w}$ is parallel with $\nabla\perron_{n_0}(\pv)$. Now, by
  \cite{Deutsch84}, we have $\frac{\partial\perron_{n_0}(\pv)}{\partial q_k}=
  y_kg'(\pve_k)\sum\nolimits_{l\in\logic} b_{k,l}^{(n_0)}x_l
  =\frac{g'(\pve_k)}{g(\pve_k)}y_k\sum\nolimits_{l\in\logic}
  g(\pve_k)b_{k,l}^{(n_0)}x_l
  =\perron_{n_0}(\pv)\frac{g'(\pve_k)}{g(\pve_k)}y_k x_k
  =\frac{g'(\pve_k)}{g(\pve_k)}y_k x_k$ for each $k\in\logic$ where $\ve{y}$
  and $\ve{x}$ are left and right positive eigenvectors of
  $\ma{G}(\pv)\ma{B}^{(n_0)}$ associated with $\perron_{n_0}(\pv)$, which, by
  irreducibility, are unique up to positive multiples.

\subsection{Proof of Corrolary \ref{cor:SpecificationWeightMaxMin}}
\label{app:proofCorSpecifWeight}

  As $\vmat$ is irreducible, Observations \ref{obs:F_Properties} and
  \ref{obs:UserPC_MaxMin_SIRBalancedUniq} (see also the proof) imply that
  $\bar{\ve{p}}$ corresponds to a point $\bar{\pv}\in\partial\freg$. Since
  $\pv^\ast(\ve{w})\in\partial\freg$ for any $\ve{w}>0$, it follows from
  Theorem \ref{th:SpecificationWeight} that $\bar{\pv}=\pv^\ast(\ve{w})$ if
  $\ve{w}$ is has the form (\ref{eq:SpecificationWeight}). Now by Observations
  \ref{obs:F_Properties} and \ref{obs:UserPC_MaxMin_SIRBalancedUniq}, we have
  $\bar{\pve}_1=\dotsb=\bar{\pve}_K$. Thus, as both $g$ and its derivative
  $g'$ are strictly monotonic (by \ref{as:UserPC_Continuous} and
  \ref{as:UserPC_LogConvex}), we must have $\ve{u}(\bar{\pv})=a\ve{1},a>0$ and
  $\ma{G}(\bar{\pv})=1/\rho(\ma{B}^{(n_0)})\ma{I}$. Thus, the corollary
  follows from Theorem \ref{th:SpecificationWeight}.

\bibliographystyle{IEEEtran}
\input{maxminSINR_stanczak.bbl}


\end{document}

%% file: makros.tex
\usepackage{courier}
\usepackage{amsmath}
\usepackage{amssymb}
\usepackage{amsxtra}
\usepackage{amsfonts}
\usepackage{stmaryrd}
\usepackage[]{graphicx}
\usepackage[dvipsnames]{color}
\usepackage{enumerate}
\usepackage[mathscr]{euscript}

\usepackage{algorithmic}
\usepackage[]{algorithm}

\newtheorem{theorem}{Theorem}
\newtheorem{corollary}{Corollary}
\newtheorem{lemma}{Lemma}
\newtheorem{definition}{Definition}
\newtheorem{remark}{Remark}
\newtheorem{observation}{Observation}

\newcommand{\field}[1]{\mathbb{#1}}
\newcommand{\menge}[1]{\mathrm{#1}}
\newcommand{\emenge}[1]{\mathscr{#1}}

\newcommand{\R}{{\field{R}}}

\newcommand{\RN}{{\field{R}}_+}
\newcommand{\RP}{{\field{R}}_{++}}

\newcommand{\pset}{{\menge{P}}}

\newcommand{\ppset}{{\menge{P}_+}}

\newcommand{\qset}{\menge{Q}}
\newcommand{\fset}{\menge{F}}
\newcommand{\freg}{\menge{F}} 
 
\newcommand{\simplex}{\menge{\Pi}_K}

\newcommand{\simplexp}{\menge{\Pi}^+_K}

\newcommand{\nodes}{\emenge{N}}

\newcommand{\logic}{\emenge{K}}

\newcommand{\operator}[1]{\mathrm{#1}}

\newcommand{\diag}{{\operator{diag}}}

\newcommand{\ma}[1]{{\boldsymbol{\mathrm{#1}}}}
\newcommand{\ve}[1]{{\boldsymbol{\mathrm{#1}}}}

\newcommand{\interior}{\mathrm{int}}

\newcommand{\sir}{\operator{SIR}}

\newcommand{\pve}{q} 
\newcommand{\pv}{\ve{q}} 

\newcommand{\perron}{\lambda}

\newcommand{\vmat}{\ma{V}} 
\newcommand{\gmat}{\ma{\Gamma}} 

\newcommand{\graph}{\mathscr{G}}

\newcounter{assume}
\def\theassume{(A.\arabic{assume})}


%% file: maxmin_problem.pstex_t
\begin{picture}(0,0)%
\includegraphics{maxmin_problem.pstex}%
\end{picture}%
\setlength{\unitlength}{3947sp}%
\begingroup\makeatletter\ifx\SetFigFont\undefined%
\gdef\SetFigFont#1#2#3#4#5{%
  \reset@font\fontsize{#1}{#2pt}%
  \fontfamily{#3}\fontseries{#4}\fontshape{#5}%
  \selectfont}%
\fi\endgroup%
\begin{picture}(7700,3157)(1517,-12415)
\put(4651,-12334){\makebox(0,0)[lb]{\smash{{\SetFigFont{14}{16.8}{\familydefault}{\mddefault}{\updefault}{\color[rgb]{0,0,0}$\sir_1$}%
}}}}
\put(8536,-12334){\makebox(0,0)[lb]{\smash{{\SetFigFont{14}{16.8}{\familydefault}{\mddefault}{\updefault}{\color[rgb]{0,0,0}$\sir_1$}%
}}}}
\put(2239,-11492){\makebox(0,0)[lb]{\smash{{\SetFigFont{14}{16.8}{\familydefault}{\mddefault}{\updefault}{\color[rgb]{0,0,0}$\ve{\gamma}$}%
}}}}
\put(1532,-9450){\makebox(0,0)[lb]{\smash{{\SetFigFont{14}{16.8}{\familydefault}{\mddefault}{\updefault}{\color[rgb]{0,0,0}$\sir_2$}%
}}}}
\put(6229,-11492){\makebox(0,0)[lb]{\smash{{\SetFigFont{14}{16.8}{\familydefault}{\mddefault}{\updefault}{\color[rgb]{0,0,0}$\ve{\gamma}$}%
}}}}
\put(5522,-9450){\makebox(0,0)[lb]{\smash{{\SetFigFont{14}{16.8}{\familydefault}{\mddefault}{\updefault}{\color[rgb]{0,0,0}$\sir_2$}%
}}}}
\put(2113,-9927){\makebox(0,0)[lb]{\smash{{\SetFigFont{14}{16.8}{\familydefault}{\mddefault}{\updefault}{\color[rgb]{0,0,0}$(\bar{\gamma}_1',\bar{\gamma}_2')$}%
}}}}
\put(3108,-10646){\makebox(0,0)[lb]{\smash{{\SetFigFont{14}{16.8}{\familydefault}{\mddefault}{\updefault}{\color[rgb]{0,0,0}$(\bar{\gamma}_1,\bar{\gamma}_2)$}%
}}}}
\put(4251,-10411){\makebox(0,0)[lb]{\smash{{\SetFigFont{14}{16.8}{\familydefault}{\mddefault}{\updefault}{\color[rgb]{0,0,0}$(\gamma_1'',\gamma_2'')$}%
}}}}
\put(6251,-9874){\makebox(0,0)[lb]{\smash{{\SetFigFont{14}{16.8}{\familydefault}{\mddefault}{\updefault}{\color[rgb]{0,0,0}$(\bar{\gamma}_1',\bar{\gamma}_2')=(\gamma_1'',\gamma_2'')$}%
}}}}
\end{picture}%

%% file: feasibleQoS.pstex_t
\begin{picture}(0,0)%
\includegraphics{feasibleQoS.pstex}%
\end{picture}%
\setlength{\unitlength}{3947sp}%
\begingroup\makeatletter\ifx\SetFigFont\undefined%
\gdef\SetFigFont#1#2#3#4#5{%
  \reset@font\fontsize{#1}{#2pt}%
  \fontfamily{#3}\fontseries{#4}\fontshape{#5}%
  \selectfont}%
\fi\endgroup%
\begin{picture}(5831,5596)(161,-12656)
\put(3376,-9661){\makebox(0,0)[lb]{\smash{{\SetFigFont{12}{14.4}{\familydefault}{\mddefault}{\updefault}{\color[rgb]{0,0,0}$\ve{w}$}%
}}}}
\put(1051,-8011){\makebox(0,0)[lb]{\smash{{\SetFigFont{12}{14.4}{\familydefault}{\mddefault}{\updefault}{\color[rgb]{0,0,0}$\rho(\ma{G}(\pv)\ma{B}^{(1)})=1$}%
}}}}
\put(2626,-11611){\makebox(0,0)[lb]{\smash{{\SetFigFont{12}{14.4}{\familydefault}{\mddefault}{\updefault}{\color[rgb]{0,0,0}$\menge{F}$}%
}}}}
\put(4426,-10786){\makebox(0,0)[lb]{\smash{{\SetFigFont{12}{14.4}{\familydefault}{\mddefault}{\updefault}{\color[rgb]{0,0,0}$\rho(\ma{G}(\pv)\ma{B}^{(2)})=1$}%
}}}}
\put(4351,-12586){\makebox(0,0)[lb]{\smash{{\SetFigFont{12}{14.4}{\familydefault}{\mddefault}{\updefault}{\color[rgb]{0,0,0}$\pve_1=\phi\bigl(\sir_1(\ve{p})/\gamma_1\bigr)$}%
}}}}
\put(901,-10861){\makebox(0,0)[lb]{\smash{{\SetFigFont{12}{14.4}{\familydefault}{\mddefault}{\updefault}{\color[rgb]{0,0,0}$\pve_1=\pve_2$}%
}}}}
\put(1576,-10036){\makebox(0,0)[lb]{\smash{{\SetFigFont{12}{14.4}{\familydefault}{\mddefault}{\updefault}{\color[rgb]{0,0,0}$\bar{\pv}=\pv^\ast(\ve{w})$}%
}}}}
\put(3834,-11706){\makebox(0,0)[lb]{\smash{{\SetFigFont{12}{14.4}{\familydefault}{\mddefault}{\updefault}{\color[rgb]{0,0,0}$\partial\menge{F}$}%
}}}}
\put(176,-7224){\makebox(0,0)[lb]{\smash{{\SetFigFont{12}{14.4}{\familydefault}{\mddefault}{\updefault}{\color[rgb]{0,0,0}$\pve_2=\phi\bigl(\sir_2(\ve{p})/\gamma_2\bigr)$}%
}}}}
\put(5926,-11324){\makebox(0,0)[lb]{\smash{{\SetFigFont{12}{14.4}{\familydefault}{\mddefault}{\updefault}{\color[rgb]{0,0,0} }%
}}}}
\end{picture}%